\documentclass[12pt, letterpaper]{article}

\usepackage[utf8]{inputenc}
\usepackage[margin=1in]{geometry}
\usepackage{amsmath, amssymb, amsfonts}
\usepackage{graphicx}
\usepackage{setspace}
\usepackage{booktabs}
\usepackage{natbib}
\usepackage{hyperref}
\hypersetup{pdftitle={Group-Level Inference with Micro-Assessments}, pdfauthor={Paul A. Jewsbury}}
\usepackage{caption}
\usepackage{subcaption}
\usepackage[capitalize]{cleveref}
\creflabelformat{equation}{#2#1#3}
\bibpunct{(}{)}{;}{a}{,}{,}
 
\title{\textbf{Group-Level Inference with Micro-Assessments}}
\author{Paul A. Jewsbury\thanks{Correspondence: Paul A. Jewsbury, Duolingo,
    5900 Penn Ave, Pittsburgh, PA 15206, USA. Email:
    \texttt{paul.jewsbury@duolingo.com}. The author wishes to thank Steven
    Nydick, Manqian (Mancy) Liao, and Alina von Davier for comments on an
    earlier draft.}\\
  Duolingo}
\date{}
 
\begin{document}
 
\maketitle

\begin{abstract}
\noindent When the target of inference is a group mean rather than an individual score, long assessments can be statistically inefficient. From the marginal score of the likelihood, we derive the asymptotic standard error of the marginal maximum likelihood estimator of a latent IRT mean under random item sampling, and obtain a closed-form approximation via a Gaussian-convolution treatment of the item information. The formula makes the test-length--sample-size trade-off explicit and includes an inflation factor for jointly estimating the population variance when the item pool is mistargeted. We validate it against exact benchmarks, Monte Carlo simulation under pool--population mismatch, and a NAEP mathematics resampling study; group means remain approximately unbiased even with single-item forms.

\end{abstract}
 
\section{Introduction}

The design of educational assessments is shaped by a tension between the demand for rigorous evidence and the constraints of instructional time. In the evaluation of educational interventions and system-wide monitoring, the primary unit of inference is rarely the individual student, but rather an aggregate unit: the classroom, the school, or the treatment arm \citep{Hedges2007, Schochet2008}. Whether validating the efficacy of a new curriculum in randomized controlled trials or monitoring proficiency trends across jurisdictions, researchers seek unbiased and precise estimates of population parameters rather than individual diagnostic classifications \citep{Mazzeo2014, Rutkowski2010}. However, the burden of testing competes directly with the instructional processes being measured, creating pressure for minimally intrusive designs \citep{Bennett2011, Wainer2000}.

Historically, this tension has been resolved inefficiently due to the conflation of individual and group-level precision. Conventional guidelines in applied research often adhere to heuristics derived from individual-specific measurement, mandating internal consistency coefficients (e.g., $\alpha > .90$; \citealp{McNeish2018, Taber2018}) that, at typical inter-item correlations of $.15$--$.20$, imply test lengths of roughly 35--50 items by the Spearman--Brown relation \citep{Nunnally1994}. While necessary for high-stakes individual diagnostics, this approach is statistically inefficient for group-level inference \citep{Hedge2018}. When the unit of analysis is the aggregate, the uncritical application of individual reliability standards results in the misallocation of scarce resources, prioritizing measurement error reduction over sampling error reduction \citep{Lord1962, Mislevy1992}.

\subsection{The Shifting Economics of Assessment and the Case for Micro-Assessments}

In many contemporary research settings, including digital learning environments, embedded assessment, and cluster-randomized trials, the cost structure of data collection is often inverted \citep{Koedinger2013, Shute2016}. Accessing large samples of test takers is increasingly feasible, but the opportunity cost of instructional time is prohibitive. This inversion is especially pronounced in large-scale digital learning platforms. A platform with tens of millions of active learners can obtain responses from very large samples at negligible marginal cost, but each administered item consumes learner time that would otherwise support instruction. The binding constraint is therefore the learner time absorbed by measurement rather than the supply of respondents. This economic shift creates an incentive for \textit{micro-assessments}: ultra-short standardized forms administered via random item sampling, in which each examinee responds to only a few items, $k$ (e.g., $k \le 10$), drawn from a larger calibrated pool.

Interest in much shorter measures is already evident across the behavioral sciences. Single-item measures have recently been the subject of an explicit call to action \citep{Allen2022}, and empirical work finds that single items can retain criterion validity comparable to their multi-item counterparts \citep{Song2023}. A single item, however, cannot span a content domain. The micro-assessment design developed here moves the content-coverage burden from the individual form to the aggregate item-sampling design, enabling content coverage with short individual forms.

Random item sampling allows researchers to decouple the length of the test from the breadth of the content domain \citep{Childs2002}. By administering unique, randomly selected subsets of items to each examinee, the design minimizes individual testing burden while maintaining a large aggregate item pool. This broad item sampling supports content representation at the aggregate level and dilutes the influence of any single aberrant item, providing robustness against item-specific misfit \citep{Rutkowski2014}. Validity still depends on the quality and representativeness of the entire item pool.

When item assignment is independent of the examinee each test form is a randomly parallel form in the sense of \cite{Lord1964}, and the framework developed here provides the group-level precision for such designs (see \citealp{Lee2024, Lee2026}, for recent treatments of individual-level properties). When item assignment depends on the examinee (e.g., targeted selection based on grade level or course placement), forms are no longer randomly parallel and overlap in the item pool across subpopulations becomes necessary to maintain a common scale \citep{Mazzeo2014}.

\subsection{Limitations of Current Random Item Sampling Designs}

The distinction between individual and group precision was recognized in the development of large-scale education surveys such as the National Assessment of Educational Progress (NAEP) and the Programme for International Student Assessment (PISA). By using matrix sampling, in which each test taker receives only a subset of the item pool \citep{Lord1962, Shoemaker1973}, these programs demonstrated that approximately unbiased group distributions could be recovered without precise individual measurement \citep{Beaton1990, Mislevy1992, OECD2012}. However, while these designs represented a significant methodological advance, they retain design constraints that make them too burdensome for many modern applications.

First, legacy large-scale assessments remain substantial batteries, typically requiring one to two hours of administration time. This duration was optimal in an era where the marginal cost of administration logistics such as recruiting schools was high, necessitating that researchers extract maximum information from every seated test taker \citep{Jones2004, Messick1983}. Second, plausible-value systems fit a latent regression on many background variables so the broad range of secondary analyses they support stays congenial with it \citep{Meng1994, vonDavier2006, Wu2005}. Because the group and population estimands are parameterized as functions of the high-dimensional coefficient vector, their estimators are functions of the many estimated coefficients, each biased away from zero by the non-linearity of the IRT model; these biases accumulate rather than cancel and grow with the number of predictors \citep{Jewsbury2025, JewsburyJohnson2025}. Direct MML instead estimates the latent mean and variance themselves, so its estimator is not assembled from many estimated quantities and incurs no such accumulation, even at single-item forms.

\subsection{Theoretical Framework: Direct Estimation and Efficiency}

The validity of micro-assessments rests on the use of \textit{direct} marginal maximum likelihood (MML) or Bayesian estimation. MML integrates over the latent distribution to estimate population parameters directly from sparse response vectors \citep{Bock1981, Mislevy1984}. As we show below, under the specified model with fixed item parameters and identifiable population moments, direct MML does not require an individual-score reliability threshold for consistency of the group mean, which holds even at $k=1$. Observed-score group differences are attenuated under low reliability \citep{Karvelis2025}, but this attenuation is a property of observed-score contrasts and does not bear on the latent mean, which MML estimates directly from the marginal likelihood.

The logic of micro-assessments aligns with the broader literature on \textit{planned missingness}. Research on planned missingness has shown that although it reduces efficiency for manifest-variable models, it remains highly efficient for latent-variable models. Specifically, \citet{Rhemtulla2016} showed that, for the structural parameters of latent-variable models, the asymptotic relative efficiency of planned-missingness designs remains high relative to the fraction of responses collected: per administered item, planned designs can outperform complete-data designs. This formalizes and extends earlier planned-missingness research (e.g., \citealp{Graham2006, Little2013}).

Despite these theoretical advances, the adoption of micro-assessments is limited by a lack of accessible design tools. While the asymptotic variance of \textit{individual} ability estimates is well established via Fisher information \citep{Lord1980}, the corresponding result for the \textit{marginal} mean estimator under random item sampling has, to our knowledge, not been expressed as a closed-form design equation.

\subsection{Current Study}

The objective of this study is to provide a statistical framework for the design of micro-assessments. Starting from the marginal score identity for the MML mean, we derive the asymptotic Fisher information for a latent group mean under random item sampling, then develop a closed-form Gaussian-convolution approximation to the resulting standard error. The approximation separates the contributions of sample size, test length, pool targeting, and joint estimation of the population variance, and we translate it into iso-precision and iso-power design contours. The framework is validated via an exact finite-length information benchmark, Monte Carlo simulation under pool--population mismatch, and an empirical application to NAEP mathematics data.

\section{Theoretical Framework}

This study derives the asymptotic sampling variance of the population mean estimator $\hat{\mu}_{\theta}$ under a dual-randomness design in which both persons and items are sampled from their respective distributions. Operating within the two-parameter logistic (2PL) framework, we distinguish between the uncertainty arising from finite sample sizes (population sampling variance) and the uncertainty arising from limited test length (measurement error).

\subsection{Statistical Model and Notation}

Let $\theta_i \in \mathbb{R}$ denote the latent ability of person $i$. The probability of a correct response ($u_{ij}=1$) by person $i$ to item $j$ is given by the 2PL model:
\begin{equation}
    P(u_{ij}=1 \mid \theta_i, a, b_{ij}) = \frac{1}{1 + \exp\left[-1.7 a (\theta_{i} - b_{ij})\right]}\label{eq:IRF}
\end{equation}
where $a$ is the item-invariant discrimination parameter, $b_{ij}$ is the difficulty of the $j$th item administered to person $i$, and $1.7$ is the conventional scaling constant.

We define the stochastic variation in the assessment design as follows:
\begin{enumerate}
    \item \textbf{Person Sampling:} Examinees are drawn i.i.d.\ from a population density $g(\theta)$ with mean $\mu_{\theta}$ and variance $\sigma^2_{\theta}$. Following the standard MML specification \citep{Bock1981, Mislevy1984}, we take $g$ to be normal, $g(\theta) = \phi(\theta; \mu_{\theta}, \sigma^2_{\theta})$, where $\phi(x; \mu, \sigma^2)$ denotes the normal probability density function.
    \item \textbf{Item Sampling:} Every examinee $i$ receives a unique form of $k$ items, with difficulties $\mathbf{b}_i = (b_{i1}, \dots, b_{ik})$ drawn i.i.d.\ from an item pool density $h(b)$ with mean $\mu_b$ and variance $\sigma_b^2$, and with a common discrimination $a$.
\end{enumerate}
Throughout, $\mu_{\theta}$ denotes the latent population mean (the focal estimand), and $\mu_b$ denotes the mean of the item pool. We further write $\Delta = \mu_b - \mu_{\theta}$ for the signed pool--population mismatch; because the design quantities below depend on $\Delta$ only through $\Delta^2$, this sign is immaterial to them, and we use $|\Delta|$ only when describing plotted magnitudes.

Let $\mathbf{u}_i = (u_{i1}, \dots, u_{ik})$ denote the response vector of person $i$. Because $\theta_i$ is not observed, inference rests on the \emph{marginal} probability of a response pattern given the administered items, obtained by integrating the conditional likelihood over the population density:
\begin{equation} \label{eq:marginal_prob}
    P(\mathbf{u}_i \mid \mathbf{b}_i) = \int_{-\infty}^{\infty} \prod_{j=1}^{k} P(u_{ij} \mid \theta, a, b_{ij})^{u_{ij}} \left[1 - P(u_{ij} \mid \theta, a, b_{ij})\right]^{1-u_{ij}} g(\theta) \, d\theta
\end{equation}
The MML estimator $(\hat{\mu}_{\theta}, \hat{\sigma}^2_{\theta})$ maximizes the marginal log-likelihood $\ell(\mu_{\theta}, \sigma^2_{\theta}) = \sum_{i=1}^{N} \ell_i$, where $\ell_i = \log P(\mathbf{u}_i \mid \mathbf{b}_i)$, the dependence on $(\mu_{\theta}, \sigma^2_{\theta})$ through $g$ left implicit \citep{Bock1981, Mislevy1984}. We emphasize that $\hat{\mu}_{\theta}$ is the maximum likelihood estimator of a structural parameter: no prior is placed on $\mu_{\theta}$ itself, and no person-level point estimates are computed.

Two features of this estimand bound the claims that follow. First, the standard errors derived below are conditional on the calibrated item parameters: $a$ and the difficulties are treated as fixed and known, so the sampling theory describes uncertainty from person and item sampling but not from item calibration, linking, or drift. This is appropriate when micro-assessments are assembled from a large, previously calibrated operational pool. When the pool is itself newly or sparsely calibrated, an additional calibration-error component is required. Second, because $g$ is specified as normal, the consistency of $\hat{\mu}_{\theta}$ concerns the model-implied latent mean: when the fitted population model matches the latent distribution this is the population mean of interest, and the reliance on $g$ is mild for long forms but maximal at $k=1$, where the population model carries nearly all the inferential weight.

\subsection{The Marginal Score and the Information for the Latent Mean}

The asymptotic behavior of $\hat{\mu}_{\theta}$ is governed by the score of the marginal likelihood. Differentiating \Cref{eq:marginal_prob} under the integral sign and using $\partial g / \partial \mu_{\theta} = g(\theta)\,(\theta - \mu_{\theta})/\sigma^2_{\theta}$ for the normal density yields a closed form for the per-person score:
\begin{equation} \label{eq:score_identity}
    \frac{\partial \ell_i}{\partial \mu_{\theta}} = \frac{\tilde{\theta}_i - \mu_{\theta}}{\sigma^2_{\theta}}, \qquad \tilde{\theta}_i \equiv E\left(\theta_i \mid \mathbf{u}_i, \mathbf{b}_i\right)
\end{equation}
where $\tilde{\theta}_i$ is the posterior mean of $\theta_i$ under the population distribution (expected a posteriori; EAP). \Cref{eq:score_identity} makes precise the role of EAP quantities in MML estimation: the score of the marginal likelihood is a functional of the EAP.

The Fisher information per examinee follows from the variance of the score, taken over both the response pattern and the random item draw:
\begin{equation} \label{eq:info_mu}
    \mathcal{I}(\mu_{\theta}) = \operatorname{Var}\left( \frac{\tilde{\theta} - \mu_{\theta}}{\sigma^2_{\theta}} \right) = \frac{\operatorname{Var}(\tilde{\theta})}{\sigma^4_{\theta}} = \frac{\sigma^2_{\theta} - E\left[\operatorname{Var}(\theta \mid \mathbf{u}, \mathbf{b})\right]}{\sigma^4_{\theta}}
\end{equation}
where the last equality is the law of total variance \citep{Casella2002}: the variance of the posterior means equals the population variance minus the expected posterior variance.

We first treat the population variance $\sigma^2_{\theta}$ as known, so that $\mu_{\theta}$ is the only free parameter. \Cref{sec:joint} relaxes this assumption and estimates $\mu_{\theta}$ and $\sigma^2_{\theta}$ jointly. By standard maximum likelihood asymptotics \citep{Cramer1946, Rao1945}, the variance of the MML mean estimator is then the reciprocal of its information,
\begin{equation} \label{eq:avar_mu}
    \operatorname{Var}(\hat{\mu}_{\theta}) \approx \frac{1}{N \, \mathcal{I}(\mu_{\theta})} = \frac{\sigma^4_{\theta}}{N \operatorname{Var}(\tilde{\theta})}
\end{equation}
Because $\operatorname{Var}(\tilde{\theta}) \le \sigma^2_{\theta}$, \Cref{eq:avar_mu} is bounded below by $\sigma^2_{\theta}/N$: no amount of testing can reduce the standard error of a group mean below the population sampling floor.

\subsubsection{Expected Information Under Random Item Sampling}

In a random item sampling design, examinees receive unique test forms drawn from an item pool density $h(b)$. The information content of the design is therefore not a sum over fixed items, but an expectation over the pool distribution. We define the design-averaged test information $\bar{I}$ as the test length multiplied by the double integral of the item information function $I(\theta \mid b)$ over both the item pool density $h(b)$ and the population density $g(\theta)$:
\begin{equation} \label{eq:exact_info_integral}
\bar{I} = k \int_{-\infty}^{\infty} \! \int_{-\infty}^{\infty} I(\theta \mid b) h(b) g(\theta) \, db \, d\theta
\end{equation}
Note that $\bar{I}$ is the expected Fisher information about $\theta$ for a single randomly assembled form, not the marginal information for $\mu_{\theta}$ (the score variance of \Cref{eq:info_mu}).

\subsubsection{Gaussian Approximation: The Additive Error Decomposition} \label{sec:ga}

\Cref{eq:avar_mu} requires the variance of the posterior means, which has no closed form. The Gaussian treatment of the response-pattern likelihood approximates the posterior precision of a given form by the prior precision plus that form's information \citep{Lord1980}. Replacing that realized information by its design expectation $\bar{I}$ is a further plug-in step. Because $x \mapsto \sigma^4_{\theta}\, x/(1 + \sigma^2_{\theta} x)$ is concave, this plug-in step overstates $\operatorname{Var}(\tilde{\theta})$ by Jensen's inequality, so the resulting closed form slightly understates the standard error. These two approximations give $E[\operatorname{Var}(\theta \mid \mathbf{u}, \mathbf{b})] \approx \sigma^2_{\theta} / (1 + \sigma^2_{\theta} \bar{I})$, hence $\operatorname{Var}(\tilde{\theta}) \approx \sigma^4_{\theta} \bar{I} / (1 + \sigma^2_{\theta} \bar{I})$, and substituting into \Cref{eq:avar_mu} yields the additive form
\begin{equation} \label{eq:se_general}
    SE(\hat{\mu}_{\theta}) \approx \sqrt{ \frac{\sigma^2_{\theta} + \sigma^2_{\epsilon}}{N} }, \qquad \sigma^2_{\epsilon} \equiv \frac{1}{\bar{I}}
\end{equation}
\Cref{eq:se_general} recovers the familiar decomposition of the error of a group mean into a population sampling component and a measurement component \citep{Lord1962, Wu2010}. With the measurement component equal to the reciprocal of the expected information, it is the reciprocal representation of the shrinkage identity in \Cref{eq:info_mu}, since $\sigma^2_{\theta} + 1/\bar{I} = \sigma^4_{\theta} / ( \sigma^2_{\theta} - E[\operatorname{Var}(\theta \mid \mathbf{u}, \mathbf{b})] )$ under the same approximation.

\subsubsection{Analytical Approximation: Gaussian Convolution}

To obtain a fully closed-form solution useful for study design, we approximate the convolution of the item information, item sampling, and population densities. The Fisher information for a 2PL item is $I(\theta \mid b) = 1.7^2 a^2 P(\theta)\left[1-P(\theta)\right]$, with $P(\theta)$ given by \Cref{eq:IRF}. The logistic function with the $1.7$ scaling constant approximates the normal ogive to a maximum absolute error below $.01$, $P(\theta) \approx \Phi\left(a(\theta - b)\right)$, where $\Phi$ is the standard normal cumulative distribution function \citep{Haley1952, Birnbaum1968, Camilli1994}. Differentiating both sides of this approximation with respect to $\theta$ gives $1.7\, a\, P(\theta)\left[1-P(\theta)\right] \approx \phi\left(\theta; b, 1/a^2\right)$, and hence:
\begin{equation} \label{eq:gauss_info}
    I(\theta \mid b) \approx (1.7 a) \cdot \phi\left(\theta; b, \frac{1}{a^2}\right)
\end{equation}

We seek the marginal expected information for a single random item, $\bar{I}_{\text{item}}$, by integrating this approximation over the population density $g(\theta) \sim N(\mu_{\theta}, \sigma^2_{\theta})$ and the item pool density $h(b) \sim N(\mu_b, \sigma^2_b)$.

First, we evaluate the inner integral with respect to $\theta$, conditional on a specific item difficulty $b$. We utilize the Gaussian product identity, which states that the integral of the product of two Gaussian PDFs over their common domain equates to a single Gaussian density function evaluated at the first mean, parameterized by the second mean and the sum of their variances \citep{Bromiley2003}:
\begin{equation}
    E_{\theta}[I(\theta \mid b)] \approx (1.7 a) \int_{-\infty}^{\infty} \phi\left(\theta; b, \frac{1}{a^2}\right) \phi\left(\theta; \mu_{\theta}, \sigma^2_{\theta}\right) d\theta
\end{equation}
\begin{equation}
    E_{\theta}[I(\theta \mid b)] \approx (1.7 a) \phi\left(b; \mu_{\theta}, \sigma^2_{\theta} + \frac{1}{a^2}\right)
\end{equation}
This intermediate result represents the expected information provided by an item of fixed difficulty $b$ when administered to a random examinee.

Next, we integrate this result over the distribution of item difficulties $h(b)$ to account for random item sampling. Applying the Gaussian product identity a second time, the variance of the item pool ($\sigma^2_b$) is added to the previous variance sum:
\begin{equation}
    \bar{I}_{\text{item}} \approx (1.7 a) \int_{-\infty}^{\infty} \phi\left(b; \mu_{\theta}, \sigma^2_{\theta} + \frac{1}{a^2}\right) \phi(b; \mu_b, \sigma_b^2) db
\end{equation}
\begin{equation} \label{eq:single_item_info}
    \bar{I}_{\text{item}} \approx (1.7 a) \phi\left(\mu_{\theta}; \mu_b, \sigma^2_{\theta} + \sigma^2_{b} + \frac{1}{a^2} \right)
\end{equation}

We define the total system variance ($\sigma^2_{sys}$) as the sum of all variance components in the measurement design:
\begin{equation}
    \sigma^2_{sys} = \sigma^2_{\theta} + \sigma^2_{b} + \frac{1}{a^2}
\end{equation}
Substituting this into \Cref{eq:single_item_info} and expanding the Gaussian density formula yields the explicit analytical function for the single-item expected information:
\begin{equation}
    \bar{I}_{\text{item}} \approx \frac{1.7 a}{\sqrt{2\pi \sigma_{sys}^2}} \exp\left( -\frac{(\mu_{\theta} - \mu_{b})^2}{2 \sigma^2_{sys}} \right)
\end{equation}

Finally, since the test consists of $k$ items drawn independently, the total expected information of \Cref{eq:exact_info_integral} factorizes as
\begin{equation} \label{eq:info_factorization}
    \bar{I} = k \, \bar{I}_{\text{item}}
\end{equation}
Substituting the inverse of this total information into the standard error equation (\Cref{eq:se_general}) provides the closed-form approximation:
\begin{equation} \label{eq:se_approx_uni}
    SE(\hat{\mu}_{\theta}) \approx \frac{1}{\sqrt{N}} \sqrt{ \sigma^2_{\theta} + \frac{1}{k} \left[ \frac{\sqrt{2\pi \sigma^2_{sys}}}{1.7 a} \exp\left( \frac{(\mu_{\theta} - \mu_{b})^2}{2 \sigma^2_{sys}} \right) \right] }
\end{equation}

The closed form assumes a common discrimination $a$, consistent with the 2PL design model above. The empirical application revisits its accuracy on a real pool with heterogeneous parameters.

\subsubsection{Joint Estimation of the Population Variance Under Pool Mismatch} \label{sec:joint}

The quantities derived so far---the additive decomposition of \Cref{eq:se_general} and the closed form in \Cref{eq:se_approx_uni}---are computed from the information for $\mu_{\theta}$ alone, and therefore presume that $\sigma^2_{\theta}$ is known or anchored. Under joint estimation the operative variance is the $(\mu_{\theta}, \mu_{\theta})$ element of the inverse of the $2 \times 2$ information matrix $\mathbf{J}$ for $(\mu_{\theta}, \sigma^2_{\theta})$, the covariance matrix of the per-person score vector $(\partial \ell_i / \partial \mu_{\theta},\, \partial \ell_i / \partial \sigma^2_{\theta})$. Writing $J_{\mu\mu} = \mathcal{I}(\mu_{\theta})$ for the mean information of \Cref{eq:info_mu} and $J_{\mu\sigma}, J_{\sigma\sigma}$ for the remaining entries, the $(1,1)$ element of the $2 \times 2$ inverse is
\begin{equation} \label{eq:joint_inverse}
    \left[ \mathbf{J}^{-1} \right]_{\mu\mu} = \frac{J_{\sigma\sigma}}{J_{\mu\mu} J_{\sigma\sigma} - J_{\mu\sigma}^2} = \frac{1}{\mathcal{I}(\mu_{\theta}) \left( 1 - \varrho^2 \right)} = \frac{C}{\mathcal{I}(\mu_{\theta})}, \qquad C \equiv \frac{1}{1 - \varrho^2},
\end{equation}
where $\varrho^2 = J_{\mu\sigma}^2 / (J_{\mu\mu} J_{\sigma\sigma})$ is the squared correlation between the two scores. The factor $C$ is a variance inflation factor in the regression sense, with the score correlation $\varrho$ in the role of a predictor correlation \citep{CoxReid1987}.

A closed-form approximation to $C$ follows from the same Gaussian-kernel treatment that produced \Cref{eq:se_approx_uni}. We state it here and derive it in Appendix~\ref{app:joint}. At $k = 1$ the marginal model is a single Bernoulli response per examinee and the kernel approximation of \Cref{eq:gauss_info} applied to its information matrix gives $\varrho^2 = \Delta^2 s^2 / (\Delta^2 s^2 + \sigma^2_b \sigma^2_{sys})$, with $s^2 = \sigma^2_{\theta} + 1/a^2$. For $k > 1$ the squared score correlation decays as $1/k$. Combining the $k = 1$ closed form with this decay gives the unified correction
\begin{equation} \label{eq:C_joint}
    C(\Delta, k) \approx 1 + \frac{\Delta^2 \left( \sigma^2_{\theta} + 1/a^2 \right)}{k \, \sigma^2_b \, \sigma^2_{sys}}
\end{equation}
and the completed design formula
\begin{equation} \label{eq:se_joint}
    SE(\hat{\mu}_{\theta}) \approx \sqrt{ \frac{\left[ \sigma^2_{\theta} + \sigma^2_{\epsilon}(\Delta, k) \right] C(\Delta, k)}{N} }, \qquad \sigma^2_{\epsilon}(\Delta, k) = \frac{1}{k \, \bar{I}_{\text{item}}}
\end{equation}
with $\bar{I}_{\text{item}}$ evaluated at the design's mismatch via \Cref{eq:single_item_info}. \Cref{eq:C_joint} is a design approximation rather than an exact $k > 1$ result: it reduces at $k = 1$ to the Gaussian-kernel closed form, equals $1$ at $\Delta = 0$ under a symmetric Gaussian pool, carries the correct $O(k^{-1})$ decay, and is numerically accurate across the design region (Appendix~\ref{app:joint}).

\section{Simulation}

To validate the analytical derivations and examine the joint effect of sample size and test length on the standard error of the latent mean, we conducted a Monte Carlo simulation study. All simulation procedures, including the data generation and the custom estimation routines described below, were implemented in the R statistical computing environment \citep{RCoreTeam}.

\subsection{Methodology}

\subsubsection{Simulation Design}
The simulation employed a fully crossed two-factor design varying sample size $N$ (100, 500, 1{,}000, 10{,}000) and test length $k$ (1, 2, 3, 4, 5, 10, 20, 40). A second grid validated the corrected design formula (\Cref{eq:se_joint}) under pool--population mismatch. Examinee abilities remained $\theta_i \sim N(0, 1)$ while the item pool mean was shifted, $b_{ij} \sim N(\Delta, 1)$, with mismatch $\Delta \in \{0, 0.5, 1, 1.5, 2\}$ crossed with test length $k \in \{1, 2, 5, 10\}$ and sample size $N \in \{250, 1{,}000\}$. For each condition, 1,500 replications were performed to ensure stable estimates of the standard error and bias.

\subsubsection{Data Generation}
Response data were generated using the 2PL model defined in \Cref{eq:IRF}. The discrimination parameter was fixed at $a = 1.0$ for all items. In the baseline grid, examinee abilities and item difficulties were both drawn from a standard normal distribution, $\theta_i \sim N(0, 1)$ and $b_{ij} \sim N(0, 1)$; in the mismatch grid the item-difficulty mean was shifted to $\Delta$ as described above, $b_{ij} \sim N(\Delta, 1)$, while $\theta_i \sim N(0,1)$ was held fixed. In each replication, per-person forms were drawn without replacement from a fresh pool of 10{,}000 difficulties, approximating i.i.d.\ sampling from $h(b)$.

\subsubsection{Estimation}

We implemented MML estimation via the expectation-maximization (EM) algorithm \citep{Bock1981}. Integration over the latent density used an equally spaced grid of $Q = 91$ quadrature nodes spanning $-9$ to $+9$, wider than a standard-normal density alone would require, so that accuracy remained uniform even at large offset. Item parameters were fixed at their generating values throughout estimation, so that only the population parameters $(\hat{\mu}_{\theta}, \hat{\sigma}^2_{\theta})$ were estimated. The EM iterations were terminated when the largest change in $(\hat{\mu}_{\theta}, \hat{\sigma}_{\theta})$ between successive steps fell below $10^{-4}$, subject to a cap of $5{,}000$ iterations, and $\hat{\sigma}^2_{\theta}$ was bounded below at $0.01$ to prevent boundary degeneracy.

\subsubsection{Outcome Measures}

To evaluate the performance of the estimation methods and derive the design trade-offs, we computed both empirical recovery statistics and analytical design contours.

\textbf{Simulation Recovery Metrics.}
For each condition, let $\mu_{true}$ denote the generating population mean and $\hat{\mu}_r$ denote the estimated mean in replication $r$, where $r = 1, \dots, R$ and $R=1,500$. We evaluated the accuracy of the group-level estimates using three metrics: bias, defined as the average deviation from the true parameter, $\text{Bias} = R^{-1} \sum_{r=1}^R (\hat{\mu}_r - \mu_{true})$; the empirical standard error (SE), defined as the standard deviation of the estimates across replications, $SE(\hat{\mu}) = \left[(R-1)^{-1} \sum_{r=1}^R (\hat{\mu}_r - \bar{\hat{\mu}})^2\right]^{1/2}$, where $\bar{\hat{\mu}}$ is the mean of the estimates; and the root mean square error (RMSE), $\text{RMSE} = \left[R^{-1} \sum_{r=1}^R (\hat{\mu}_r - \mu_{true})^2\right]^{1/2}$, which combines both bias and variance into a single measure of overall accuracy. Monte Carlo uncertainty in these metrics is summarized by $95\%$ intervals (chi-square for the SE, normal for the bias, delta method for the RMSE).

\textbf{Calculation of Iso-Design Curves.}
Trade-off curves (iso-precision and iso-power contours; see \citealp{Baker2021}) were calculated analytically by inverting the per-examinee variance equation. We defined the total variance per examinee as $\sigma^2_{obs}(k) = \sigma^2_{\theta} + \sigma^2_{\epsilon}(k)$, with the measurement-error component $\sigma^2_{\epsilon}(k) = 1/\bar{I}$ obtained from the closed-form Gaussian convolution (\Cref{eq:single_item_info,eq:info_factorization}). The contours assume a targeted pool ($\Delta = 0$, so $C = 1$). For off-target designs the per-examinee variance becomes $[\sigma^2_{\theta} + \sigma^2_{\epsilon}(\Delta, k)]\,C(\Delta, k)$ (\Cref{eq:se_joint}).

For the \textbf{iso-precision} curves, the required sample size $N$ for a target standard error $SE_{target}$ was solved as:
\begin{equation}
    N = \frac{\sigma^2_{obs}(k)}{SE_{target}^2}
\end{equation}

For the \textbf{iso-power} curves, we solved for $N$ iteratively. The required sample size was determined by finding the root of the power function for the non-central $t$-distribution. We test $H_0\!: \delta = 0$ against a two-sided alternative at level $\alpha$ for a treatment effect $\delta = \mu_{exp} - \mu_{con}$, so that the sample size satisfies
\begin{equation}
    \left[1 - T_{df}(t_{1-\alpha/2,\,df} \mid \lambda)\right] + T_{df}(-t_{1-\alpha/2,\,df} \mid \lambda) = 1 - \beta,
\end{equation}
where $T_{df}(\cdot \mid \lambda)$ is the noncentral $t$ cumulative distribution function with noncentrality $\lambda = \delta / SE(\hat{\delta})$, $1 - \beta = 0.80$, and $\alpha = 0.05$. We compute the power contours under a large-benchmark comparison, in which the control mean is known with negligible error (e.g., contrasting a cohort with a state norm), and under a balanced two-arm design with equal sample size per arm. The degrees of freedom follow the design: $df = N - 1$ for the one-sample benchmark comparison and $df = 2N - 2$ for the balanced two-arm design. Because $\hat{\mu}_{\theta}$ is asymptotically normal, the non-central $t$ reference distribution \citep{Johnson1995} is a finite-sample convention accounting for the estimation of $\sigma^2_{obs}$ that converges to the normal approximation as the degrees of freedom grow. For off-target pools $\sigma^2_{obs}(k)$ is replaced by the corrected variance $[\sigma^2_{\theta} + \sigma^2_{\epsilon}(\Delta,k)]\,C(\Delta,k)$ of \Cref{eq:se_joint}.

\subsubsection{Calculation of Theoretical Benchmarks}
Empirical results are compared against two quantities. The first is the closed-form design formula---the Gaussian convolution (\Cref{eq:se_approx_uni}) for the centered grid, and its corrected form (\Cref{eq:se_joint}) for the mismatch grid. The second is an exact information benchmark obtained without the Gaussian-kernel approximation, computed by the number-correct reduction described in Appendix~\ref{app:joint}.

\subsection{Results}

\subsubsection{Convergence of Analytical and Empirical Precision}

\begin{figure}[htbp]
    \centering
    \includegraphics[width=0.9\textwidth]{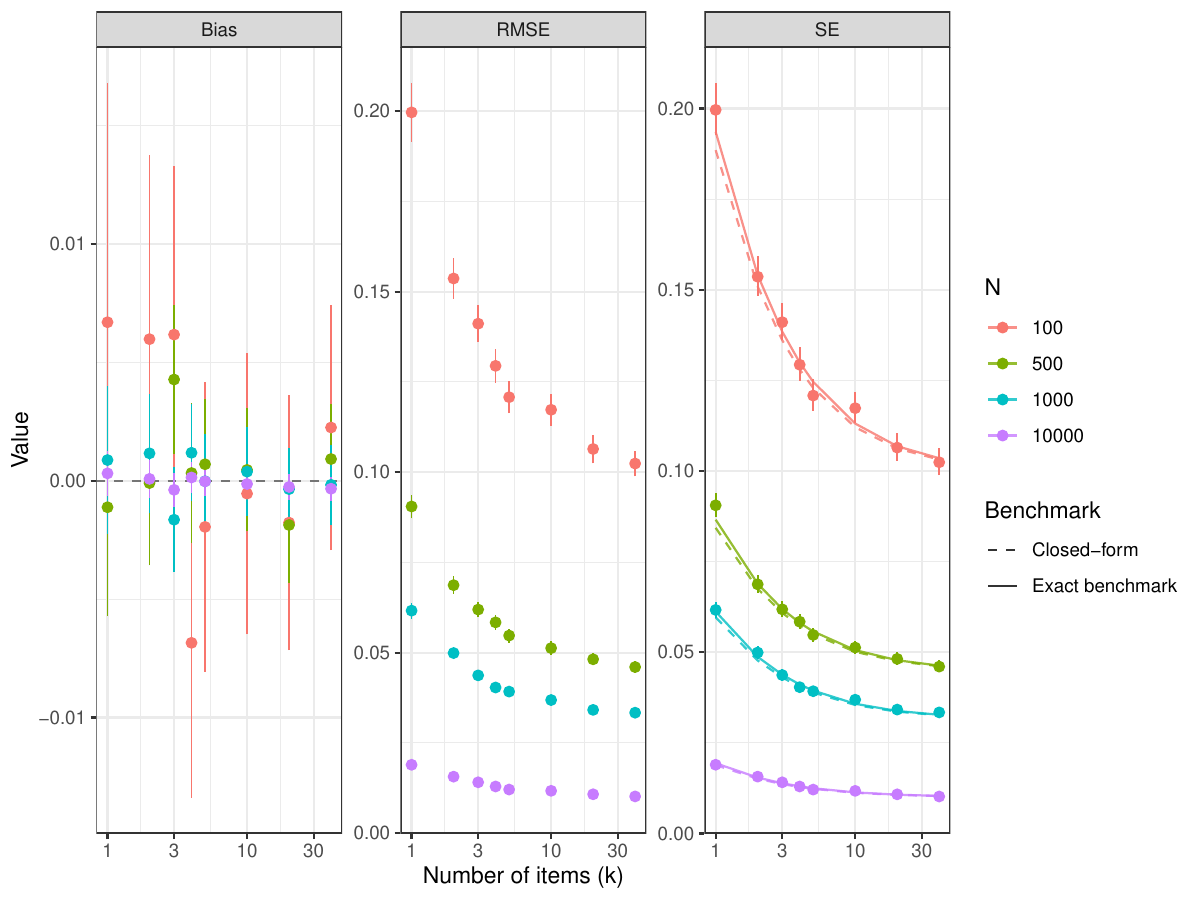}
    \caption{Validation under a targeted pool ($\Delta = 0$). Points show empirical bias, SE, and RMSE across varying $N$ and $k$; lines on the SE panel show the closed-form design formula (\Cref{eq:se_approx_uni}) and the exact marginal-information benchmark. Vertical bars are Monte Carlo $95\%$ intervals (chi-square for the SE, normal $\pm 1.96\,\widehat{SE}/\sqrt{R}$ for the bias, and a delta-method interval for the RMSE; $R = 1{,}500$).}
    \label{fig:uni_val}
\end{figure}

Figure \ref{fig:uni_val} presents the baseline results under a targeted item pool ($\Delta = 0$, so the inflation factor $C = 1$). Three primary patterns emerge from the analysis. First, no evidence of estimator bias was observed in any condition. The bias was negligible even for designs with a single item response per test taker ($k=1$). Consequently, the RMSE was driven almost entirely by the standard error.

Second, the results substantiate the theoretical dominance of sample size ($N$) over test length ($k$) for group-level inference. While increasing test length reduced the standard error, the benefit showed steep diminishing returns; improvements in precision were negligible beyond $k=10$. In contrast, sample size acted as the primary constraint on precision. For instance, a design with $N=1,000$ and $k=1$ consistently yielded a smaller standard error than a design with $N=100$ and $k=40$.

Finally, the closed-form expression closely tracked both the empirical standard errors and the exact marginal-information benchmark. The exact benchmark lies above the closed form by 2.7\%, 2.1\%, 1.4\%, and 0.9\% on the standard-error scale at $k = 1, 2, 5,$ and $10$, respectively. The empirical standard errors exceeded both curves only at $k = 1$, consistent with the exact bound acting as a finite-sample information floor that the maximum likelihood estimator attains only asymptotically and approaches from above as $N$ grows.

\subsubsection{Validation Under Pool--Population Mismatch}

\begin{figure}[htbp]
    \centering
    \includegraphics[width=1.0\textwidth]{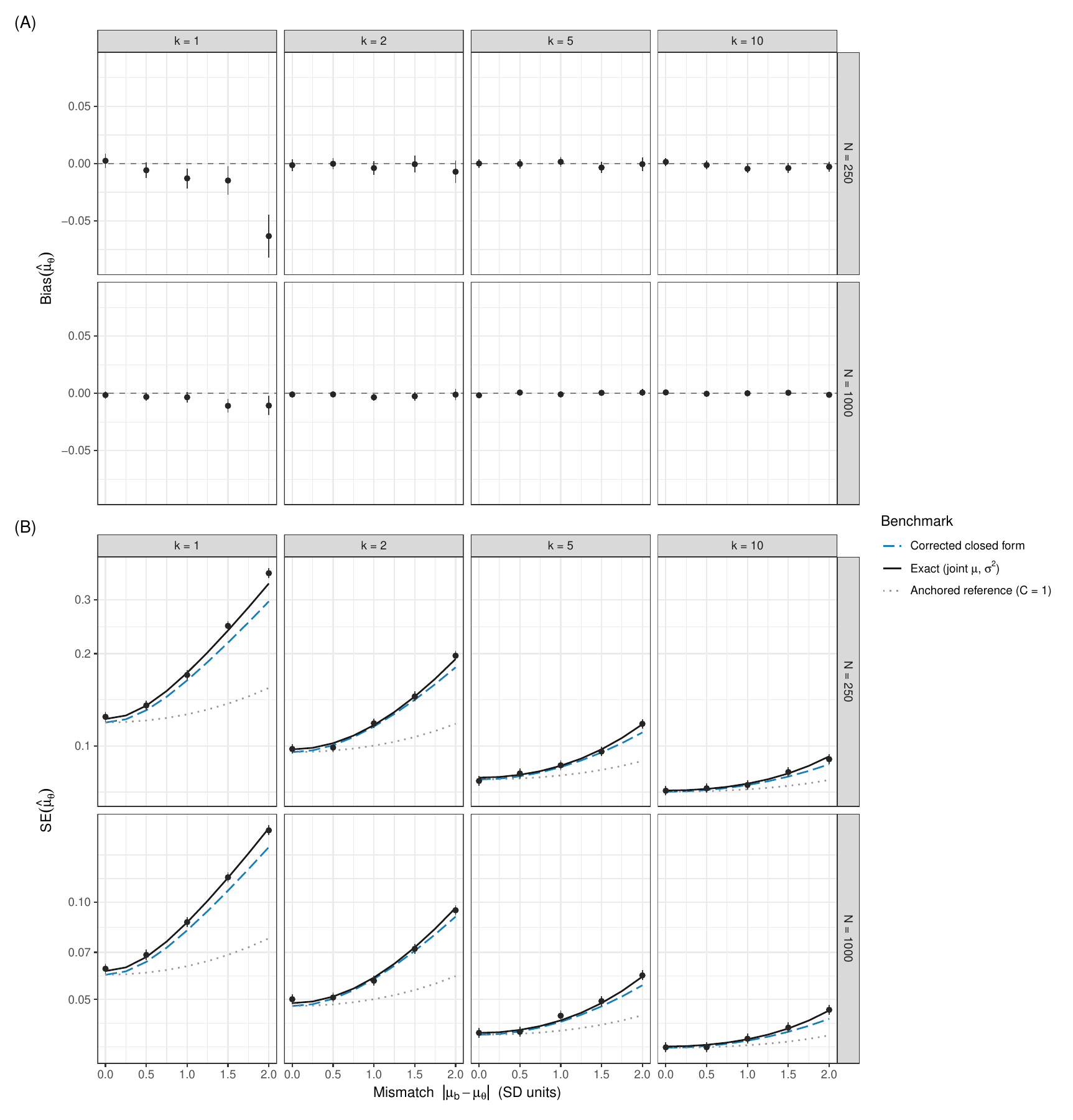}
\caption{Validation under pool--population mismatch, by test length ($k$, columns) and sample size ($N$, rows). \textbf{(A)} Empirical bias of $\hat{\mu}_{\theta}$ on a linear axis with a shared, symmetric range; the dashed line marks zero, the target value, and points carry Monte Carlo intervals (no benchmark line is needed, as zero is the reference). \textbf{(B)} Standard error of $\hat{\mu}_{\theta}$ as a function of the mismatch magnitude $|\Delta| = |\mu_b - \mu_{\theta}|$: lines show the corrected closed form (\Cref{eq:se_joint}), the exact joint-information benchmark (Appendix~\ref{app:joint}), and the anchored-$\sigma^2_{\theta}$ reference ($C = 1$), and points show empirical standard errors with Monte Carlo intervals. In (B) the vertical axis is logarithmic and shared across test lengths within each sample-size row, so the decline of $SE(\hat{\mu}_{\theta})$ with $k$ is directly comparable across columns. Both blocks share the $\Delta$ axis.}    \label{fig:mismatch}
\end{figure}

Figure~\ref{fig:mismatch} presents the mismatch grid. Bias in $\hat{\mu}_{\theta}$ was approximately zero and within Monte Carlo noise at every test length beyond a single item (Figure~\ref{fig:mismatch}A). Only at $k = 1$ did bias drift from zero as the mismatch grew, reaching at most $0.04$ in the most demanding condition ($\Delta = 2$, $N = 250$). All cells converged under the estimation budget, and the $\hat{\sigma}^2_{\theta}$ floor bound in only 1 of 60{,}000 replications.

The empirical standard errors track the exact joint-information benchmark across all conditions: over the 40 mismatch cells, the ratio of the empirical standard error to the benchmark fell within $0.97$--$1.03$ (Monte Carlo standard error $\approx 1.8\%$ per cell), except at the most demanding cell ($k = 1$, $\Delta = 2$, $N = 250$), where finite-sample variance exceeded the asymptotic floor (ratio $1.08$, falling to $0.99$ at $N = 1{,}000$). The corrected closed form (\Cref{eq:se_joint}) is accurate to within roughly $2.5\%$ on the standard-error scale for $k \ge 2$ and $9\%$ at $k = 1$ for $\Delta \le 1.5$, and understates by $6\%$ ($k \ge 2$) to $13\%$ ($k = 1$) at $\Delta = 2$ (Appendix~\ref{app:joint} reports the full range to $\Delta = 3$).

The two specifications isolate the cost of estimating $\mu_{\theta}$ and $\sigma^2_{\theta}$ jointly. The standard error grows with mismatch in both, but faster under joint estimation, so it increasingly exceeds the anchored-$\sigma^2_{\theta}$ reference ($C = 1$) as $\Delta$ grows. This gap is the inflation $C(\Delta, k)$ of \Cref{eq:C_joint}.

\subsubsection{Design Trade-offs}
\begin{figure}[htbp]
    \centering
    \includegraphics[width=1.0\textwidth]{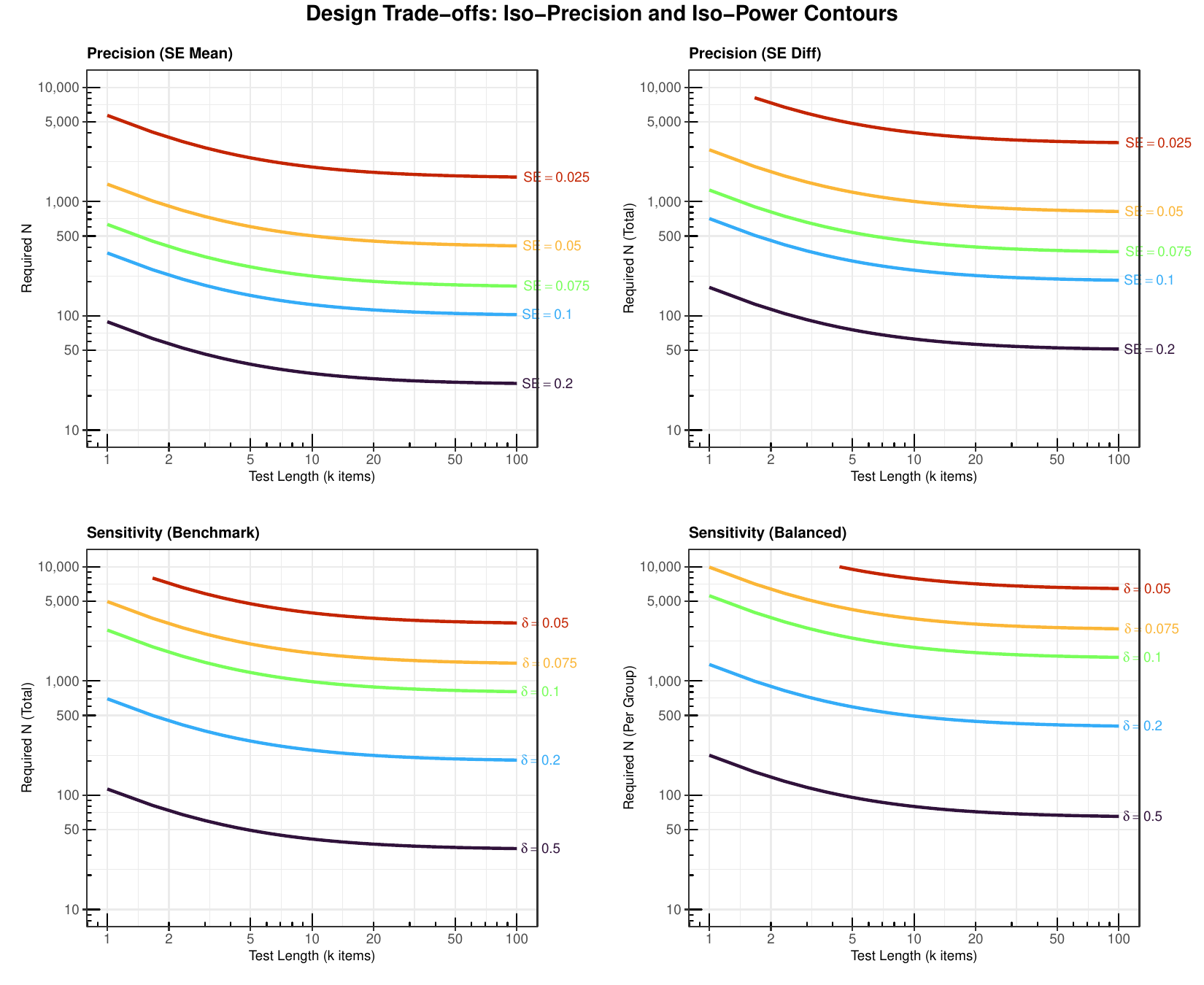}
    \caption{Design Trade-off Contours. Top panels show sample sizes required for fixed precision; lower panels show sample sizes required for fixed power ($1-\beta=0.80$). In the lower panels, $N$ is the experimental-arm size for the large-benchmark comparison and the per-arm size for the balanced two-arm design. Contours assume a targeted pool ($\Delta = 0$, $C = 1$); off-target, the per-examinee variance becomes $[\sigma^2_{\theta} + \sigma^2_{\epsilon}(\Delta, k)]\,C(\Delta, k)$ (\Cref{eq:se_joint}), so the required $N$ rises through both the mismatch penalty in $\sigma^2_{\epsilon}$ and the joint-estimation factor $C(\Delta, k)$.}
    \label{fig:tradeoffs}
\end{figure}

Following the power contour framework of \citet{Baker2021}, we present iso-precision and iso-power contours in Figure \ref{fig:tradeoffs}, illustrating the indifference points between sample size ($N$) and test length ($k$).

The top panels in Figure \ref{fig:tradeoffs} confirm that the total error variance is dominated by population sampling variance rather than measurement error for all but the shortest tests. While increasing test length from $k=1$ to $k=10$ yields noticeable efficiency gains by reducing the instrument noise, these benefits rapidly diminish. For instance, achieving a standard error of 0.05 is feasible with a 10-item test ($N \approx 500$) or a 100-item test ($N \approx 400$). This comparison highlights the inefficiency of traditional long forms for group-level inference: a ten-fold increase in testing burden reduces the required sample size by only 20\%, confirming that resources are more efficiently allocated to increasing $N$ once a minimal length is achieved.

The lower panels in Figure \ref{fig:tradeoffs} extend this logic to statistical power. The contours demonstrate that rigorous hypothesis testing is viable even with ultra-short forms, provided the sample size is adjusted to compensate for lower reliability. For example, under the large-benchmark design, detecting a small effect size ($\delta=0.2$) is achievable with as few as 5 items given an experimental-arm sample of $N \approx 300$; the corresponding balanced two-arm design requires approximately twice this number in each arm. While increasing length to $k=20$ reduces this requirement to $N \approx 220$, the contours illustrate that $k$ and $N$ are functional substitutes. Consequently, researchers can trade test length for sample size to minimize instructional disruption without sacrificing model-based precision or power, provided the calibrated item pool and population model remain appropriate.

\section{Empirical Application: NAEP Mathematics Assessment}

To validate the analytical derivations under realistic conditions, we conducted a resampling study with empirical data from the 2005 Grade 8 National Assessment of Educational Progress (NAEP) mathematics assessment, administered by the U.S. National Center for Education Statistics (NCES) and accessed through the public-use sample distributed with the \texttt{NAEPprimer} package \citep{NAEPprimer}. Whereas the simulations used synthetic 2PL data, this application uses a three-parameter logistic (3PL) framework, accommodating guessing, empirical item-parameter distributions, and model misfit.

The empirical analysis was implemented in the R statistical computing environment \citep{RCoreTeam}. IRT model parameters were estimated with the \texttt{mirt} package \citep{Chalmers2012}.

\subsection{Methodology}

\subsubsection{Data and Calibration}
The Primer student file contains 17{,}606 records (assessed and excluded students from the public-school sample); we analyzed the $N = 16{,}915$ students in the reporting sample and retained the $N = 9{,}542$ with at least 30 valid responses, avoiding sparsity-related artifacts unrelated to test design. The pool comprised the $J = 143$ cognitive items listed in the Primer's item-parameter tables, fitting a constrained unidimensional 3PL with pseudo-guessing fixed at 0.20 for dichotomous items with nonzero NAEP guessing parameters (i.e., multiple-choice items) and at 0 otherwise. Polytomous items were dichotomized by scoring any partial or full credit as correct (special missing codes treated as missing) to align the application with the dichotomous theory of \Cref{eq:IRF}; the application is thus a stress test of the design approximation rather than an official NAEP calibration. Treating the filtered response matrix as the population, we obtained a global MML calibration and held its item parameters fixed throughout. Full-data benchmark means and variances were then estimated separately for the Overall, Gender (Male, Female), and Race/Ethnicity (White, Black, Hispanic) groups, each by a single-group model with the calibrated item parameters held fixed and only the latent mean and variance freed. Group pool sizes after filtering were 9{,}542 (Overall), 4{,}742 (Male), 4{,}800 (Female), 5{,}907 (White), 1{,}625 (Black), and 1{,}323 (Hispanic); the remaining 687 students in other race/ethnicity categories were not analyzed separately. The analysis is unweighted and conditions on the filtered analytic sample, providing internal validation against a full-data IRT benchmark rather than official NAEP population estimates.

\subsubsection{Resampling Design}
We used a fully crossed design over demographic group $G$ (Overall, Male, Female, White, Black, Hispanic), test length $k$ (1, 2, 3, 5, 10, 20, 30), and sample size $N$ (250, 500, 1{,}000), with $R = 1{,}500$ replications per condition. Each replication drew $N$ students with replacement from the relevant group pool (so that replication variability estimates superpopulation sampling variability), reusing the same student draw across test lengths within a replication, and, for each student, constructed a length-$k$ form by sampling $k$ responses without replacement from their observed response vector.  Because each student's available items reflect the original NAEP assessment design, this procedure thins observed responses rather than implementing a fully prospective design in which any student could receive any of the 143 items.

\subsubsection{Estimation and Outcome Measures}
Group means were estimated with unidimensional single-group 3PL models in which the full-data item parameters were held fixed and only the latent moments freed; each replication's group model was initialized at the full-sample estimates. We evaluated the performance of the micro-assessments using two metrics: bias, the average deviation of the estimated mean from the full-data benchmark mean across replications, $E[\hat{\mu} - \mu_{\text{bench}}]$, and the empirical standard error (SE), the standard deviation of the estimates across replications, $SD(\hat{\mu})$.

\subsection{Results}

\subsubsection{Unbiasedness}
\begin{figure}[htbp]
    \centering
    \includegraphics[width=1.0\textwidth]{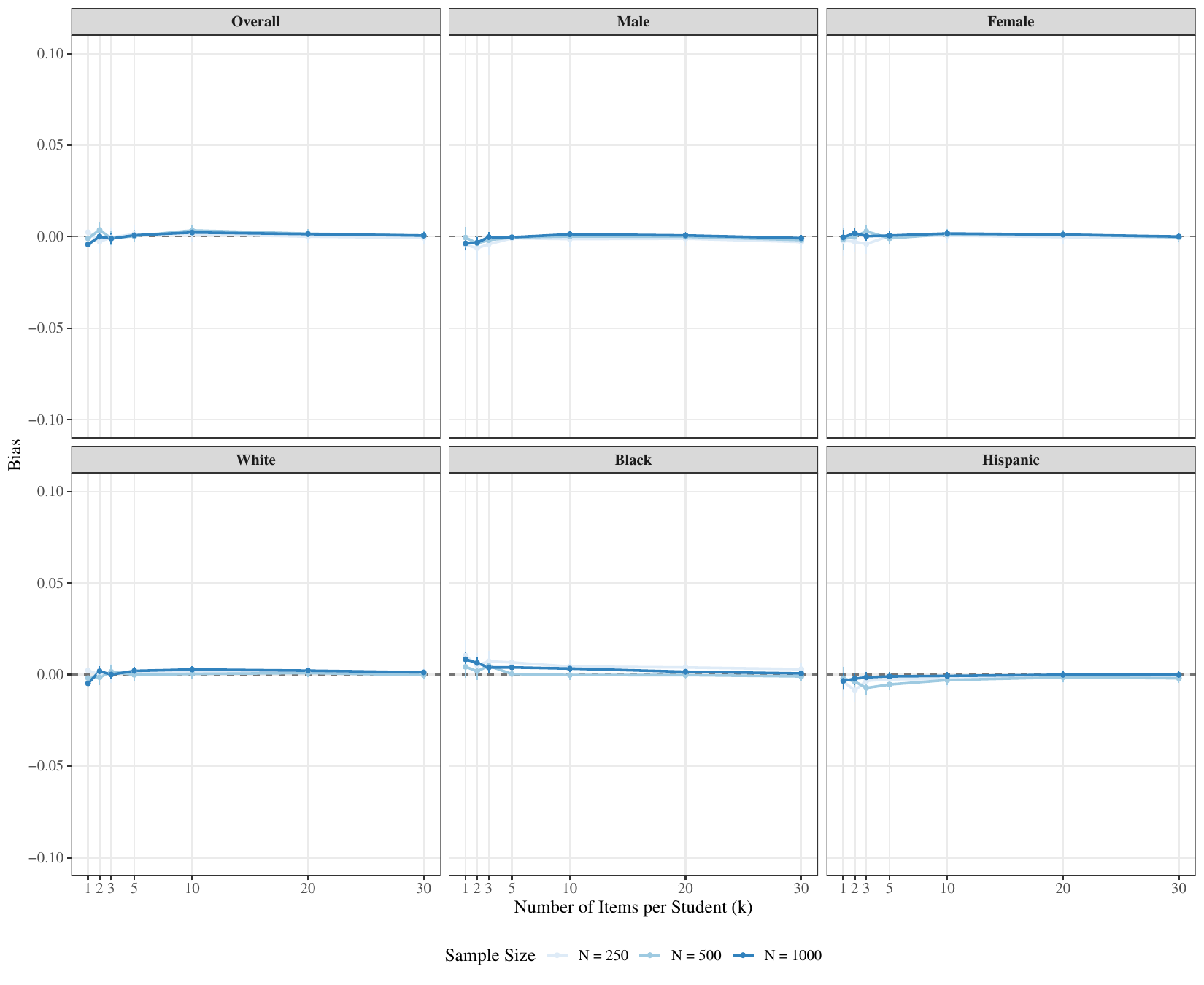}
    \caption{\textbf{Empirical Bias of Group Mean Estimates.} Bias is calculated as the mean difference between the micro-assessment estimate and the full-data benchmark mean computed from the complete response matrix. Vertical bars are Monte Carlo $95\%$ intervals for the bias ($\pm 1.96\,\widehat{SE}/\sqrt{R}$, with $R$ the per-condition number of successfully fitted replications); intervals covering zero are consistent with unbiased estimation.}
    \label{fig:naep_bias}
\end{figure}

Consistent with the simulation findings, the MML mean estimator remained approximately unbiased across all test lengths and sample sizes (Figure~\ref{fig:naep_bias}), and across all demographic subgroups regardless of proficiency distribution. The estimator agreed with the full-data benchmark without systematic error even at $k = 1$, indicating that integration over the latent distribution allows single-item sparsity without introducing bias. All $189{,}000$ fits ($126$ conditions $\times$ $1{,}500$ replications) returned estimates. Bias never exceeded $0.010$ in absolute value (at most $0.06$ of the corresponding standard error, at $k = 1$, $N = 250$ for Black students), with Monte Carlo half-widths of comparable size.

\subsubsection{Precision Trade-offs (N vs. k)}
\begin{figure}[htbp]
    \centering
    \includegraphics[width=1.0\textwidth]{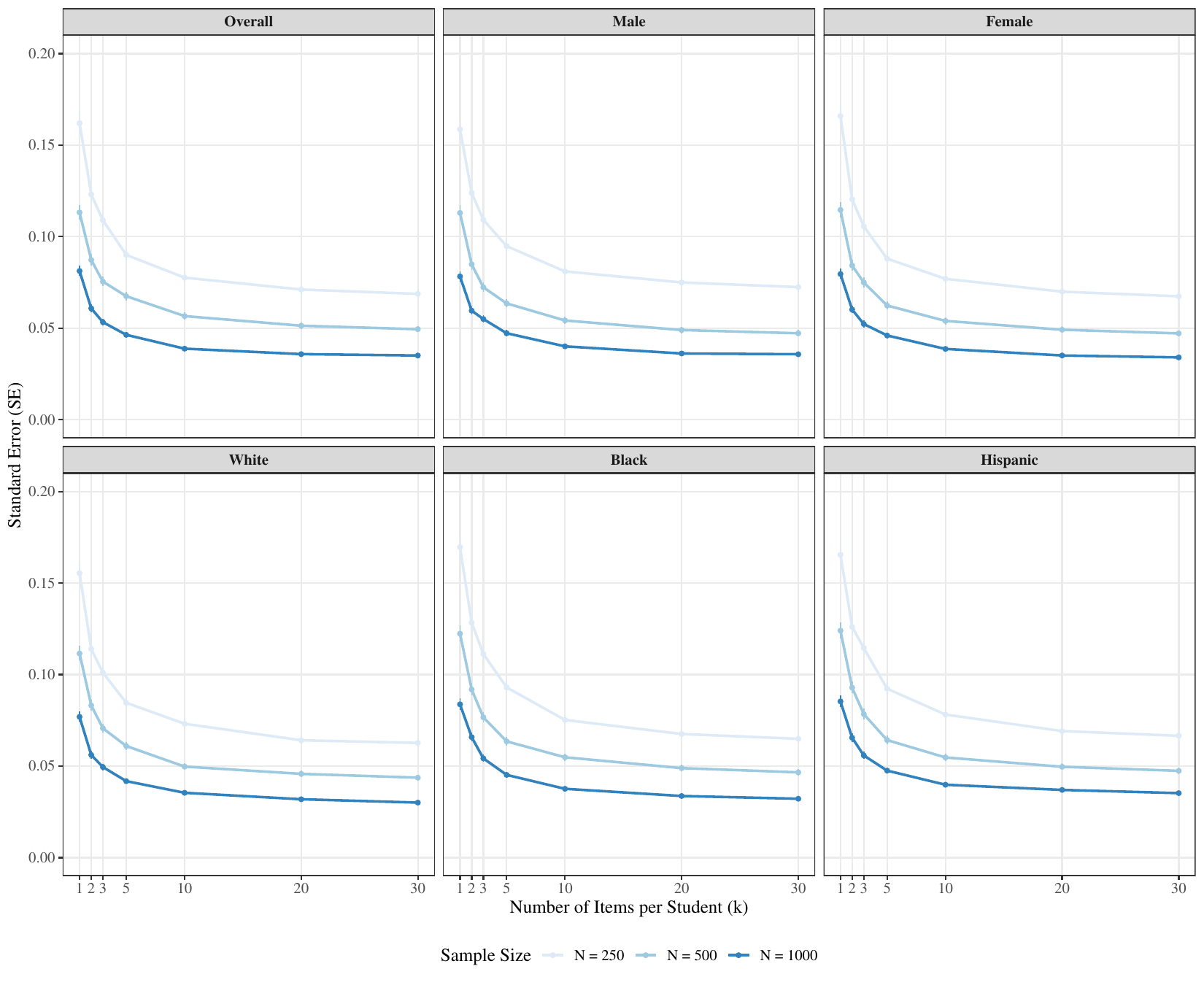}
    \caption{\textbf{Empirical Precision Trade-offs.} The standard error (SE) of the mean estimator is plotted as a function of test length ($k$) for varying sample sizes ($N$). Vertical bars are Monte Carlo $95\%$ intervals for the empirical SE (chi-square, based on the per-condition number of successfully fitted replications).}
    \label{fig:naep_se}
\end{figure}

Figure \ref{fig:naep_se} illustrates the precision (standard error) of the group mean estimates across subgroups, empirically demonstrating the trade-off between sample size and test length.

The results replicate the scaling law of \Cref{eq:se_general}: test length yields diminishing returns while larger samples shift the entire error curve downward. Subgroups with means further from the pool center (Black and Hispanic students) showed standard errors 6--12\% above those of White students at matched $N$ and $k$, consistent with the targeting penalty of \Cref{eq:se_approx_uni} compounded by the joint-estimation inflation of \Cref{eq:C_joint}, since each subgroup's moments were estimated jointly. Fitting the per-examinee variance form $N \cdot \widehat{SE}^2 = \hat{\sigma}^2_{\theta} + \hat{B}/k$ within each group localizes this penalty in the measurement component: the implied per-item information is $0.154$ (Black) and $0.157$ (Hispanic) versus $0.194$ (White), the empirical counterpart of the targeting factor $\exp\{-\Delta^2/(2\sigma^2_{sys})\}$ in \Cref{eq:single_item_info}. The compensatory relationship follows the same $1/k$ structure: at $N = 250$, roughly nine items were required to match the precision of a single-item form at $N = 1{,}000$ (group-specific values between 6.5 and 9.8); equivalently, the five-item form at $N = 250$ carried an 8--21\% larger standard error across groups.

\section{Discussion}

These results challenge the assumption that long, highly reliable tests are a prerequisite for rigorous educational research. When the primary target of inference is a group parameter---such as a mean proficiency for a grade level or an effect size in an efficacy trial---the reliability of the individual score is secondary to the precision of the aggregate estimator \citep{Mislevy1992, Rutkowski2010, Wu2010}. This study formalizes the efficiency trade-offs inherent in micro-assessments, providing an analytical framework for optimizing precision under strict constraints on instructional time.

\subsection{Sampling Variance Dominates Measurement Error}

A primary implication of our derivation is that the conventional heuristic of maximizing internal consistency (e.g., $\alpha > .90$; \citealp{McNeish2018, Taber2018}) leads to the misallocation of resources in group-level studies. The standard error of the group mean is governed by two additive sources of uncertainty: population sampling variance and measurement error variance. Consequently, the total variance of the group mean is asymptotically bounded from below by the sampling variance of the population. Once the measurement error becomes negligible relative to this sampling variance (approximately $k=10$ well-targeted dichotomous items; Figure~\ref{fig:uni_val}), the marginal utility of additional items approaches zero. At this point, resources are better allocated to increasing sample size, which linearly attenuates the sum of both variance components.

This finding supports the growing demand for minimally intrusive measurement in learning. Rather than administering infrequent, intrusive batteries, systems can use frequent micro-assessments to monitor system-level performance without disrupting instruction \citep{Bennett2011, Koedinger2013}. This shift from ``audit'' testing to continuous, low-stakes monitoring allows for more responsive educational interventions while maintaining rigorous aggregate precision \citep{Shute2016}.

\subsection{The Compensatory Design of Micro-Assessments}

Our findings demonstrate a robust compensatory relationship between sample size and test length. Under the fitted model, fixed item parameters, and the conditions studied here, MML estimation remained approximately unbiased for the group mean down to a single item per student. This implies researchers can trade instrument precision for sample size without biasing the group-mean estimate.

As shown in our design contours (Figure \ref{fig:tradeoffs}), a micro-assessment with $k=1$ can achieve the same standard error as a long-form test, provided the sample size is increased to compensate for the higher noise per observation. This flexibility allows for the design of protocols where the burden on any single student is negligible, yet the aggregate inference remains rigorous. This trade-off is particularly advantageous in large-scale program evaluations where recruiting larger samples is often more feasible and less costly than securing extended testing time for each subject \citep{Kraft2020}.

\subsection{Extensions to Longitudinal and Targeted Designs}

While this study focused on cross-sectional estimation, the framework extends naturally to longitudinal monitoring. Measuring growth across grades introduces a tension between targeting efficiency and vertical scale maintenance. The penalty for mismatch between item difficulty and population proficiency is doubly multiplicative: the measurement-error variance grows exponentially in the squared mismatch (\Cref{eq:se_approx_uni}) and, under joint estimation of the population variance, the total variance is further inflated (\Cref{eq:C_joint}). The exponential penalty motivates targeted item sampling, in which item-selection probabilities depend on observed covariates such as grade so that subsets of a single common pool are probabilistically directed to specific cohorts \citep{Chang1996}. The probabilistic overlap in item selection between adjacent cohorts naturally conforms to a nonequivalent groups with anchor test (NEAT) linking design \citep{Kolen2014}. Furthermore, when micro-assessments are administered repeatedly, individual-level growth parameters can achieve adequate reliability by aggregating across time points, even when point-in-time measures are sparse \citep{Raudenbush2002, Willett1989}.

\subsection{Direct Estimation and Measurement Error in Group-Level Estimands}

A growing body of work shows that how item responses are reduced to scores can distort group- and treatment-level estimands. When responses are collapsed into sum scores or plugged-in point estimates and then analyzed, measurement error propagates into the target parameter: sum scores bias latent growth trajectories and their variance components \citep{KuhfeldSoland2022}, and scoring and calibration choices can materially change the magnitude of estimated treatment effects \citep{Soland2024}. A common solution is a post hoc errors-in-variables correction applied to fitted scores \citep{LockwoodMcCaffrey2014}. Direct MML avoids the need for any post hoc correction by integrating over the latent density (\Cref{eq:score_identity}). A second, closely related literature isolates the contribution of \emph{item sampling} to the uncertainty of aggregate effects. Treating the administered items as a sample from a domain, \citet{GilbertKimMiratrix2023} and \citet{GilbertItemHTE2025} show that ignoring item-level heterogeneity yields standard errors that are too small, while \citet{GilbertVAM2025} show, via generalizability theory, that value-added models conditioning on the realized items overstate reliability and overestimate differences between units. Our framework is the design-side complement to this analysis-side diagnosis: by deriving the standard error of $\hat{\mu}_{\theta}$ under random item sampling (\Cref{eq:se_general,eq:se_approx_uni}), the item-sampling variance is built into the design calculation from the outset.

\subsection{Assumptions and Threats to Validity} \label{sec:assumptions}

The validity of micro-assessment inference rests on assumptions that our validation studies only partially test. The consistency of $\hat{\mu}_{\theta}$ presumes the fitted population density $g$ matches the latent density. With long forms the data dominate $g$ in the posterior of \Cref{eq:score_identity}, but at $k = 1$ the reliance on $g$ is maximal, so the single-item unbiasedness we observe should be read as recovery of the model-implied latent mean under a correctly specified population model rather than as a distribution-free guarantee. Our validation studies do not fully probe this dependence: the simulation generates $\theta$ from the same normal family that the estimator fits, and the NAEP benchmark is itself a full-data normal-MML estimate, so both evaluate agreement with the long-form MML estimand under a common specification rather than recovery under a skewed, heavy-tailed, or otherwise nonnormal latent distribution. Establishing the behavior of single-item forms under such departures is an important direction we leave to future work. Relatedly, fixing item parameters from a parent calibration assumes invariance across administration contexts: items calibrated in hour-long ordered booklets may function differently delivered individually \citep{Debeer2013}, and, as noted in the theory, the standard errors condition on those parameters and so omit calibration and linking uncertainty. Finally, micro-assessments spread data thinly: at $k = 1$, $N = 1{,}000$ over a 143-item pool, each item accrues about seven responses, adequate for ability mean estimation but too few for empirical evaluation of item-specific properties.

\subsection{Conclusion}

By shifting the analytical focus from individual diagnostic classification to population parameter estimation, researchers can substantially reduce the testing burden placed on test takers. This paper provides the mathematical tools---closed-form standard error approximations, including the mismatch and joint-estimation corrections, with their associated precision and power design contours---to design short, efficient assessments that respect instructional time while retaining the statistical power required for educational policy research and longitudinal monitoring \citep{Jewsbury2023, Kraft2020}.

\onehalfspacing

\doublespacing

\appendix
\section{Joint Information and the Inflation Factor} \label{app:joint}

This appendix derives and validates the inflation factor $C(\Delta, k)$ of \Cref{eq:C_joint}.

\subsection*{Exact joint information via the number-correct score}
The score for $\sigma^2_{\theta}$ follows from differentiating the marginal likelihood with respect to the variance of the normal population density:
\begin{equation} \label{eq:score_sigma}
    \frac{\partial \ell_i}{\partial \sigma^2_{\theta}} = \frac{\left( \tilde{\theta}_i - \mu_{\theta} \right)^2 + \operatorname{Var}\left( \theta_i \mid \mathbf{u}_i, \mathbf{b}_i \right) - \sigma^2_{\theta}}{2 \sigma^4_{\theta}}
\end{equation}
which, like the score for $\mu_{\theta}$ in \Cref{eq:score_identity}, has mean zero at the true parameters for every realized item set. Both scores depend on the response pattern only through the posterior mean $\tilde{\theta}_i$ and the posterior variance $\operatorname{Var}(\theta_i \mid \mathbf{u}_i, \mathbf{b}_i)$, which, because the items share a common discrimination, depend on $\mathbf{u}_i$ only through the number-correct score $r_i = \sum_j u_{ij}$. For a realized difficulty vector the $2 \times 2$ joint information matrix $\mathbf{J}(\mathbf{b})$, the expected outer product of the score vector $(\partial \ell / \partial \mu_{\theta},\, \partial \ell / \partial \sigma^2_{\theta})$ over the marginal pattern distribution, therefore reduces exactly from a sum over the $2^k$ patterns to a $(k+1)$-term sum over $r_i \in \{0, \dots, k\}$ weighted by the marginal score probabilities $P(r_i \mid \mathbf{b}_i)$, obtained by integrating the Poisson--binomial distribution of $r_i$ given $\theta$ (Lord--Wingersky recursion, \citealp{LordWingersky1984}; $O(k^2)$, hence tractable at any test length) against $g(\theta)$. The design-level matrix $\bar{\mathbf{J}}$ is the expectation of $\mathbf{J}(\mathbf{b})$ over the pool density: deterministic quadrature at $k \le 2$, and Monte Carlo over $10^{5}$ item draws with common random numbers across $\Delta$ otherwise (Monte Carlo standard errors of the inflation $\le 0.005$).

\subsection*{Gaussian-kernel closed form at $k = 1$}
At $k = 1$ the marginal model is exactly one Bernoulli response with success curve $\bar{P}(b) = E_{\theta}\left[ P(\theta, b) \right] \approx \Phi(\eta)$, where $\eta = (\mu_{\theta} - b)/s$ and $s^2 = \sigma^2_{\theta} + 1/a^2$. Differentiating $\bar{P}$ with respect to $(\mu_{\theta}, \sigma^2_{\theta})$ gives the per-item information matrix
\begin{equation}
    \mathbf{J}(b) = w(\eta) \, \mathbf{z} \mathbf{z}^{\top}, \qquad \mathbf{z} = \left( \frac{1}{s}, \; -\frac{\eta}{2 s^2} \right)^{\!\top}, \qquad w(\eta) = \frac{\bar{P}'(\eta)^2}{\bar{P}(\eta)\left[ 1 - \bar{P}(\eta) \right]}
\end{equation}
The matrix is rank one because a single binary response is one degree of freedom; full-rank design-level information requires variation in $\eta$, that is, $\sigma^2_b > 0$. Treating the marginal curve as a $1.7$-scaled logistic in $\eta$ with unit discrimination, the kernel approximation of \Cref{eq:gauss_info} gives $w(\eta) \approx 1.7\, \phi(\eta; 0, 1)$. Under $b \sim N(\mu_b, \sigma^2_b)$ the standardized design points are $\eta \sim N(\eta_0, \tau^2)$ with $\eta_0 = -\Delta/s$ and $\tau^2 = \sigma^2_b / s^2$, and the Gaussian product identity gives the three design-level entries a common factor:
\begin{equation}
    E[w] \propto \phi\!\left( \eta_0;\, 0,\, 1 + \tau^2 \right), \qquad E[\eta\, w] = E[w] \cdot m, \qquad E[\eta^2 w] = E[w] \left( m^2 + v \right)
\end{equation}
with product mean $m = \eta_0 / (1 + \tau^2)$ and variance $v = \tau^2 / (1 + \tau^2)$. The squared information correlation is therefore $\varrho^2 = m^2 / (m^2 + v)$, which simplifies to
\begin{equation} \label{eq:rho_closed}
    \varrho^2 = \frac{\Delta^2 s^2}{\Delta^2 s^2 + \sigma^2_b \, \sigma^2_{sys}}, \qquad C_1(\Delta) = \frac{1}{1 - \varrho^2} = 1 + \frac{\Delta^2 s^2}{\sigma^2_b \, \sigma^2_{sys}}
\end{equation}
Both limits are informative: $C_1(0) = 1$ recovers the symmetric case, and, for $\Delta \neq 0$, $C_1 \to \infty$ as $\sigma^2_b \to 0$ recovers the $k = 1$ identification failure of a constant-difficulty design. This pole is specific to the single-item case: for $k > 1$, repeated responses identify $\sigma^2_{\theta}$ even at $\sigma^2_b = 0$, so the limit lies outside the intended domain of the closed form.

\subsection*{Test-length scaling}
Expanding the exact scores to leading order shows the cross-information accruing per item while the information for $\sigma^2_{\theta}$ accrues with the square of the total information, so $\varrho^2 = O(1/k)$ in the low-information regime $\sigma^2_{\theta}\, k \bar{I} \lesssim 1$ characteristic of micro-assessments. Writing $A = \Delta^2 s^2 / (\sigma^2_b \sigma^2_{sys})$ for the single-item value $C_1 - 1$ of \Cref{eq:rho_closed}, the simplest functional form consistent with both the kernel $k = 1$ value and this $O(1/k)$ decay is $\varrho^2_k = A/(k + A)$, which yields $C(\Delta, k) = 1/(1 - \varrho^2_k) = 1 + A/k$---precisely \Cref{eq:C_joint}, and distinct from a naive $\varrho^2_1/k$ rescaling for $k > 1$ (the two agree only at $k = 1$). The $k > 1$ expression is therefore a calibrated design approximation rather than an exact result, validated against the exact joint information below.

\subsection*{Validation}

\begin{center}
\small
\setlength{\tabcolsep}{5pt}
\renewcommand{\arraystretch}{0.95}
\captionof{table}{Exact joint-estimation inflation $[\bar{\mathbf{J}}^{-1}]_{\mu\mu} \bar{J}_{\mu\mu}$ versus the closed form $C(\Delta, k)$ of \Cref{eq:C_joint} ($a = 1$, $b \sim N(\Delta, 1)$, $\theta \sim N(0, 1)$). Each cell: exact / closed form. Exact values: quadrature (deterministic) at $k \le 2$; Monte Carlo over $10^{5}$ item draws (MC standard errors $\le 0.005$) at $k = 5, 10$.}
\label{tab:joint_validation}
\begin{tabular}{ccccc}
\toprule
$\Delta$ & $k = 1$ & $k = 2$ & $k = 5$ & $k = 10$ \\
\midrule
0.5 & 1.18 / 1.17 & 1.06 / 1.08 & 1.02 / 1.03 & 1.01 / 1.02 \\
1.0 & 1.72 / 1.67 & 1.27 / 1.33 & 1.09 / 1.13 & 1.04 / 1.07 \\
1.5 & 2.70 / 2.50 & 1.66 / 1.75 & 1.24 / 1.30 & 1.11 / 1.15 \\
2.0 & 4.22 / 3.67 & 2.32 / 2.33 & 1.52 / 1.53 & 1.26 / 1.27 \\
2.5 & 6.46 / 5.17 & 3.35 / 3.08 & 1.99 / 1.83 & 1.53 / 1.42 \\
3.0 & 9.76 / 7.00 & 4.95 / 4.00 & 2.76 / 2.20 & 2.01 / 1.60 \\
\bottomrule
\end{tabular}
\end{center}

\noindent Table~\ref{tab:joint_validation} compares \Cref{eq:C_joint} with the exact inflation for the simulation design ($a = 1$, $\sigma_{\theta} = \sigma_b = 1$, so $C = 1 + \tfrac{2}{3}\Delta^2/k$). On the inflation-factor scale the closed form tracks the exact benchmark within $4\%$ for $\Delta \le 1.5$ and $7\%$ at $\Delta = 2$ (slightly overstating it for $k \ge 2$), with understatement growing to $10$--$15\%$ at $\Delta = 3$ (outside the empirical grid), where the information correlation saturates toward one. On the standard-error scale of \Cref{eq:se_joint}, the corrected closed form is accurate to within roughly $2.5\%$ for $k \ge 2$ and $9\%$ at $k = 1$ for $\Delta \le 1.5$, and understates by $6$--$13\%$ at $\Delta = 2$ and $17$--$21\%$ at $\Delta = 3$. By contrast, omitting the correction understates the standard error by $54$--$70\%$ at $\Delta \ge 2$ for single-item forms.

\end{document}